\documentclass[]{spie}  

\usepackage{amsmath,amsfonts,amssymb}
\usepackage{graphicx}
\usepackage[colorlinks=true, allcolors=blue]{hyperref}
\usepackage{lipsum}
\usepackage{mwe}
\usepackage{upgreek}
\usepackage{tabularx}
\usepackage{caption}

\title{The scientific payload of the Ultraviolet Transient Astronomy Satellite (ULTRASAT)}

\author{Sagi Ben-Ami}
\author[a]{Yossi Shvartzvald}
\author[a]{Eli Waxman}
\author[a]{Udi Netzer}
\author[a]{Yoram Yaniv}
\author[b]{Viktor M. Algranatti}
\author[a]{Avishay Gal-Yam}
\author[a]{Ofer Lapid}
\author[a]{Eran Ofek}
\author[a]{Jeremy Topaz}
\author[c]{Iair Arcavi}
\author[d]{Arooj Asif} 
\author[e]{Shlomi Azaria}
\author[e]{Eran Bahalul}
\author[d]{Merlin F. Barschke}  
\author[d]{Benjamin Bastian-Querner} 
\author[d,f]{David Berge} 
\author[d]{Vlad D. Berlea}  
\author[d]{Rolf B\"uhler} 
\author[d]{Louise Dittmar} 
\author[e]{Anatoly Gelman}
\author[d]{Gianluca Giavitto}  
\author[a]{Or Guttman}
\author[d]{Juan M. Haces Crespo}  
\author[e]{Daniel Heilbrunn}
\author[e]{Arik Kachergincky}
\author[d]{Nirmal Kaipachery} 
\author[d,f]{Marek Kowalski} 
\author[d]{Shrinivasrao R. Kulkarni} 
\author[d]{Shashank Kumar} 
\author[h]{Daniel K\"usters} 
\author[a]{Tuvia Liran}
\author[e]{Yonit Miron-Salomon}
\author[e]{Zohar Mor}
\author[e]{Aharon Nir}
\author[e]{Gadi Nitzan}
\author[d]{Sebastian Philipp} 
\author[d]{Andrea Porelli} 
\author[e]{Ilan Sagiv}
\author[d,f]{Julian Schliwinski} 
\author[e]{Tuvia Sprecher}
\author[d]{Nicola De Simone} 
\author[e]{Nir Stern}
\author[g]{Nicholas C. Stone} 
\author[c]{Benny Trakhtenbrot}
\author[d]{Mikhail Vasilev} 
\author[d]{Jason J. Watson} 
\author[d]{Steven Worm} 
\author[d]{Francesco Zappon} 

\affil[a]{Department of Particle Physics and Astrophysics, Weizmann Institute of Science, Herzl St 234, Rehovot, Israel}
\affil[b]{Israel Space Agency, Tel Aviv, Israel}
\affil[c]{School of Physics and Astronomy, Tel Aviv University, Tel Aviv, Israel}
\affil[d]{Deutsches Elektronen-Synchrotron DESY, Platanenallee 6, 15738 Zeuthen, Germany}
\affil[e]{ElOp - Elbit Systems Ltd., Rehovot, Israel}
\affil[f]{Institut f\"ur Physik, Humboldt-Universit\"at zu Berlin, Newtonstrasse 15, 12489 Berlin, Germany}
\affil[g]{Racah Institute of Physics, The Hebrew University, Jerusalem, 91904, Israel}
\affil[h]{Department of Physics, University of California at Berkeley, 366 LeConte Hall MC 7300, Berkeley, CA, 94720-7300}

\authorinfo{Correspondence Author: S.Ben-Ami, sagi.ben-ami@weizmann.ac.il, Telephone: +972-8-9342067\\ }

\begin{document} 

\maketitle

\begin{abstract}
The Ultraviolet Transient Astronomy Satellite (ULTRASAT) is a space-borne near UV telescope with an unprecedented large field of view ($200\,$deg$^2$). The mission, led by the Weizmann Institute of Science and the Israel Space Agency in collaboration with DESY (Helmholtz association, Germany) and NASA (USA), is fully funded and expected to be launched to a geostationary transfer orbit in Q2/Q3 of 2025. With a grasp 300 times larger than GALEX, the most sensitive UV satellite to date, ULTRASAT will revolutionize our understanding of the hot transient universe, as well as of flaring galactic sources. We describe the mission payload, the optical design and the choice of materials allowing us to achieve a point spread function of $\sim10\,$arcsec across the FoV, and the detector assembly. We detail the mitigation techniques implemented to suppress out-of-band flux and reduce stray light, detector properties including measured quantum efficiency of scout (prototype) detectors, and expected performance (limiting magnitude) for various objects.
\end{abstract}


\section{INTRODUCTION}
\label{sec:intro}  
ULTRASAT is a scientific satellite carrying a near UV optimized (NUV; $230-290\,$nm) telescope with an exceptionally wide field of view (FoV) of $200\,$deg$^2$. It is led by the Weizmann Institute of Science and the Israel space agency in collaboration with DESY (Helmholtz association, Germany) and NASA (USA), and planned to be launched to geostationary transfer orbit (GTO), with operation at geostationary orbit (GEO) commencing at Q2/Q3 2025. The mission is targeting time-domain astrophysics phenomena, and is the first wide FoV telescope operating in the NUV band. For hot sources (T$\sim20,000\,$K) ULTRASAT is comparable in grasp, the amount of volume of space probed per unit time\cite{Ofek2020}, to that of the Vera C. Rubin observatory\cite{LSST}, the largest ground-based optical transient survey planned to begin operation in 2024. 

ULTRASAT will undertake the first wide-field UV time-domain sky survey. It will explore a new parameter space in energy (NUV), and time-scale (6 month stares, with a 5-minute continuous cadence). For target of opportunity triggers (ToOs), ULTRASAT will be able to slew in minutes to $>50\%$ of the sky. Some of the key scientific questions and phenomena ULTRASAT will target, and in which it is expected to have a significant impact, are: 
\begin{itemize}
    \item The study of gravitational wave sources: The discovery of electro-magnetic emission following the detection of gravitational waves (GW) from the mergers of binaries involving neutron stars\cite{2016ARNPS..66...23F,2017ApJ...848L..32M}. Such detections will be the key to using these events for addressing fundamental physics questions, such as the origin of the heaviest elements and the expansion rate of the universe\cite{2018MNRAS.481.3423W}. For GW triggers, ULTRASAT's wide field-of-view amply covers the angular error regions expected to be provided by GW detectors starting 2025\cite{2018LRR....21....3A}. It will provide continuous UV light curves as well as early alerts that will enable ground-based follow-up spectroscopy and monitoring of optical and infrared emission predicted to arise later. 
    \item Early detection of Supernovae: ULTRASAT will collect early UV light curves of hundreds of core-collapse supernovae (CC-SNe) to measure the radii and surface composition of their massive progenitors, as well as to determine explosion parameters\cite{2016ApJ...820...57G}. Connecting the pre-explosion stars with their diverse explosive output will chart how the population of massive stars impact their environment through mass loss and explosion, and will specify initial conditions for explosion models\cite{2011ApJ...728...63R,2017ApJ...848....8R,2020ApJ...899...51S}. Mass loss tracers will further constrain pre-explosion evolution, allowing comprehensive investigation of the final evolution and explosive death of massive stars. During its 3 years mission, ULTRASAT is expected to detect $>40$ CC-SNe during the shock breakout phase, and $>500$ CC-SNe during the shock cooling phase.
    \item Determine the high energy flare frequency distribution of stars and guide future exoplanet atmospheric studies: ULTRASAT will monitor $\sim10^6$ stars. The long baseline high cadence survey strategy will allow us to characterize UV flares and determine the low frequency - high energy tail of the flaring distribution function, specifically from dM stars, attractive exoplanet hosts for atmospheric characterization due to favorable radii ratio, known to be prodigiously UV flaring stellar objects \cite{MUSCLES,HAZMAT}. ULTRASAT will measure, for the first time, the NUV flare frequency and luminosity distribution for stars as functions of both spectral subclass, and stellar rotation period. This will allow the community to asses which stellar types are favored as habitable planet hosts and guide future spectroscopic bio-marker searches with, \textit{e.g.}, JWST and the upcoming ELTs \cite{Snellen,Ben-Ami}. 
    \item Active galactic nuclei: ULTRASAT is poised to provide novel insights into the central engines of Active Galactic Nuclei (AGN), which are powered by accreting supermassive black holes (SMBHs), as the emission from these systems peaks in the UV and is known to vary on essentially all timescales. The study of AGN-related phenomena with ULTRASAT is expected to be broadly split into two regimes. First, ULTRASAT will monitor samples of persistent, vigorously accreting SMBHs over unprecedently short timescales, much shorter than the dynamical timescale in the innermost parts of their accretion disks\cite{2018ApJ...864...27S}. This will advance our understanding of persistent accretion disks\cite{2015ApJ...806..129E}, and may provide new ways to quantify key SMBH properties, such as mass, growth rate, and even spin\cite{2013ApJ...779..187K}. Second, ULTRASAT will discover and survey SMBH-related, UV-bright transient phenomena, marking extreme changes to the accretion flows and allowing us to see how SMBHs "turn on" or "off" on exceedingly short timescales\cite{2019ApJ...883...94T,2019NatAs...3..242T}, and to obtain insights regarding super-Eddington accretion\cite{2017ApJ...843..106B,2019MNRAS.489..524K}.
    \item Tidal Disruption Events: Stars torn apart in tidal disruption events (TDEs) emit luminous ultraviolet flares\cite{2011blho.book..286G,2014ApJ...793...38A,2020SSRv..216..124V}. ULTRASAT will be able to detect hundreds to thousands of events per year (of which $\sim50$ will be at optical magnitudes brighter than 19 and thus easy to follow up from the ground). These discoveries will allow us to constrain the currently debated emission mechanisms of TDEs \cite{2020SSRv..216..114R}, measure their rates as a function of redshift \cite{2016MNRAS.461..371K} and probe their peculiar host galaxy preferences \cite{2014ApJ...793...38A,2016ApJ...818L..21F}.
\end{itemize}
All ULTRASAT data will be transmitted to the ground in real-time, and transient alerts will be distributed to the community in $<15\,$minutes, driving vigorous ground-based follow-up of static, variable and transient ULTRASAT sources. ULTRASAT is planned for a 3-year operation at a GEO orbit, with fuel sufficient to enable an extension to 6 years of operation. It is planned to reach space prior to the early phase of the planned full sensitivity operation of the GW detector network. In the following paper we describe the ULTRASAT scientific payload design, derived performance, and unique solutions and mitigation techniques adopted for such a unique instrument. The payload prime contractor is ElOp - Elbit Systems (Isreal), sensors design and manufactured by Analog Value (Israel) and Tower Semiconductors (Israel) respectively, with detector assembly contributed by DESY (Germany) - all managed by the mission program office at the Weizmann Institute of Science in Israel. The spacecraft is developed and manufactured by Israel Aircraft Industries (IAI). An overview of the mission and its science goals will be publishes in Shvartzvald et al. (in prep.), and further detailed discussion of the payload will be given in future publications by the authors.

\section{Payload overview}
\label{sec:payload}  
\begin{figure}
\begin{center}
\begin{tabular}{cc}
\includegraphics[height=6.5cm]{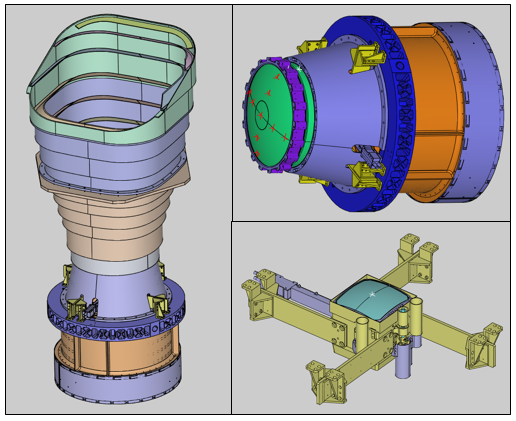} &
\includegraphics[height=6.5cm]{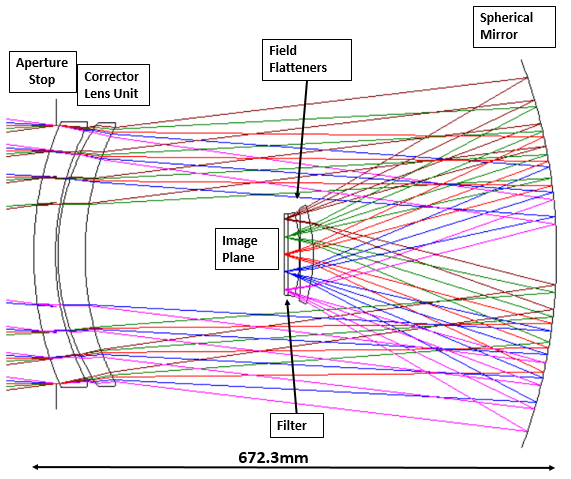}
\end{tabular}
\end{center}
\caption 
{ \label{Fig:ULTRASAT_Layout}
The ULTRASAT payload layout. Left: CAD model of the Payload, the optical tube assembly and the detector assembly; Right: Raytrace layout. The system is a catadioptric one loosely based on a Schmidt design.} 
\end{figure}

We partition the payload to three parts: Optical tube assembly (OTA), detector assembly (DA) and baffle, see Figure \ref{Fig:ULTRASAT_Layout} left panel. Each part is described below, with detailed information presented in future publications.

\subsection{Optical Tube Assembly}
The ULTRASAT payload, whose optical layout is shown in Figure \ref{Fig:ULTRASAT_Layout} right panel, is a wide FoV catadioptric system, loosely based on a Schmidt design\cite{Schmidt}. First order parameters are detailed in Table \ref{tab:ULTRASAT_optics}. Following a large ($\sim800\,$mm in length) baffle, the beam is refracted by two aspheric meniscus corrector lenses, a fused silica and a CaF$_2$ one, defining the system aperture of $330\,$mm. While the two substrates are not achromatic pairs in the classical sense at the operation bandpass, the limited choice of high transmission materials in the UV dictates their use as such. The refracted beam is then reflected by a spherical mirror, which delivers most of the optical power of the system. While in a standard Schmidt design, the correctors are placed at the mirror center of curvature, limits on the length of the payload requires us to place the correctors $\sim100\,$mm closer to the mirror - which is compensated by adding power, 1.136 diopters, to the corrector lenses. The reflected beam is then brought to a focus following refraction by a CaF$_2$ and a fused silica field flatteners. The last optical element in front of the detector array is a Sapphire filter with custom coating for out of band suppression, see section \ref{sec:OoB}. 

\begin{table}[]
    \centering
    \begin{tabular}{|l|c|c|}
        \hline
 Property & Value & Comment \\ [0.5ex] 
 \hline\hline
        Design &  Modified Schmidt& \\ \hline
        Number of powered elements & Five  & 4 refractive, 1 reflective \\ \hline
        Aperture & $330\,$mm & \\ \hline
        Design Waveband & $230-290\,$nm  & \\ \hline
        Field of View & $204\,$deg$^2$ & optimized for central $170\,$deg$^2$  \\ \hline
        Focal length & $360\,$mm & \\ \hline
        F/\# & $1.09$ & \\ \hline
        Plate Scale & $0.57\,$arcsec$\,\mu$m$^{-1}$ & \\ \hline
        Throughput & Peak $\sim30\%$  & \\ \hline
    \end{tabular}
    \caption{First order parameters of the ULTRASAT telescope.}
    \label{tab:ULTRASAT_optics}
\end{table}

\begin{figure}
\begin{center}
\begin{tabular}{cc}
\includegraphics[height=5cm]{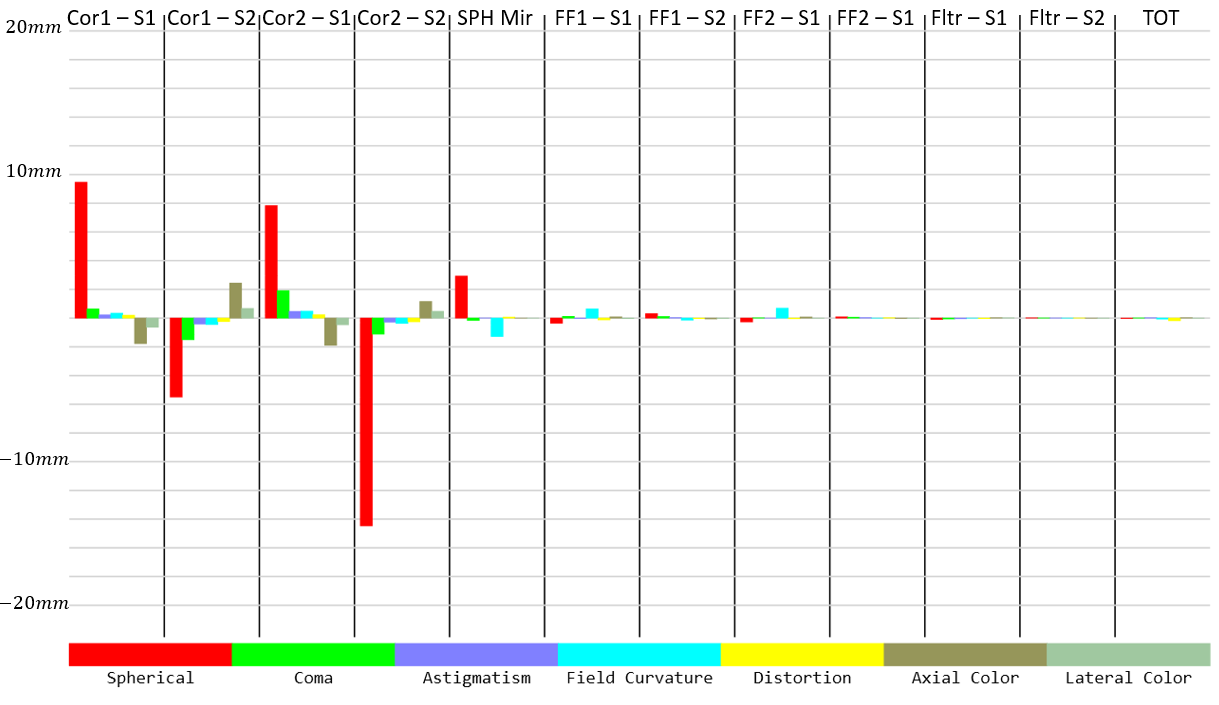} &
\includegraphics[height=5.2cm]{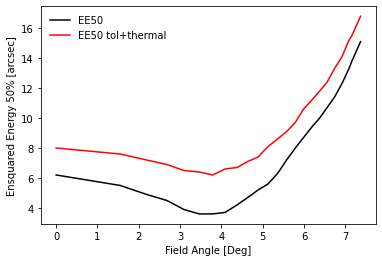}
\end{tabular}
\end{center}
\caption{Left: Seidel diagram of the ULTRASAT telescope. Corrector lenses balance spherical aberrations from the mirror, while field curvature is balanced by the two field flatteners; Right: Ensqaured energy (50\%). Nominal performance are shown in black. Red curve shows expected performance when taking manufacturing and assembly tolerances, as well as thermal gradients on the external corrector and field flattenr, into account ($90\%$ CL).} 
\label{Fig:ULTRASAT_Performance}
\end{figure}

The optical design minimizes the Seidel aberrations as shown in Figure \ref{Fig:ULTRASAT_Performance} left panel, with the corrector lenses canceling 3rd order spherical aberration. The choice of meniscus correctors also allows us to balance aberrations between on-axis and off-axis field points. Field flatteners positioned $5\,$mm from the focal plane  allows us to control field curvature. The nominal $50\%$ mean encapsulated energy diameter (EE50) across the central $170\,$deg$^2$ of the FoV is $7.5\,$arcsec, falling to $10.2\,$arcsec when taking into account manufacturing and assembly tolerances ($90\%$ CL), as well as thermal gradient uncertainties on the external lens - as the telescope is kept at a temperature of $\sim21^{\circ}$C while corrector lens is radiating to open space - see Figure \ref{Fig:ULTRASAT_Performance} right panel. While the former are compensated by serial manufacturing and through a flexible assembly program that allows correction of \textit{e.g.}, element tilt of the field flattener unit, a focus mechanism will compensate for the latter by translating one field flattener lens with respect to the other. The required resolution of the focus mechanism is $10\,\mu$m. All optical elements are coated with custom anti reflection coatings (ARC) to minimize Fresnel losses.

\subsection{Detector Assembly}
In the following section we summarise the design of the focal plane array (FPA) and the detector assembly (DA). Further information regarding the detector design can be found in proceeding $12181-109$ of this volume ('The design of a UV CMOS sensor for the ULTRASAT space telescope', T. Liran et al.) and on an earlier design stage of the DA in a previous publication\cite{2021SPIE11821E..0UA}. 

\begin{figure}
\begin{center}
\begin{tabular}{c}
\includegraphics[height=9cm]{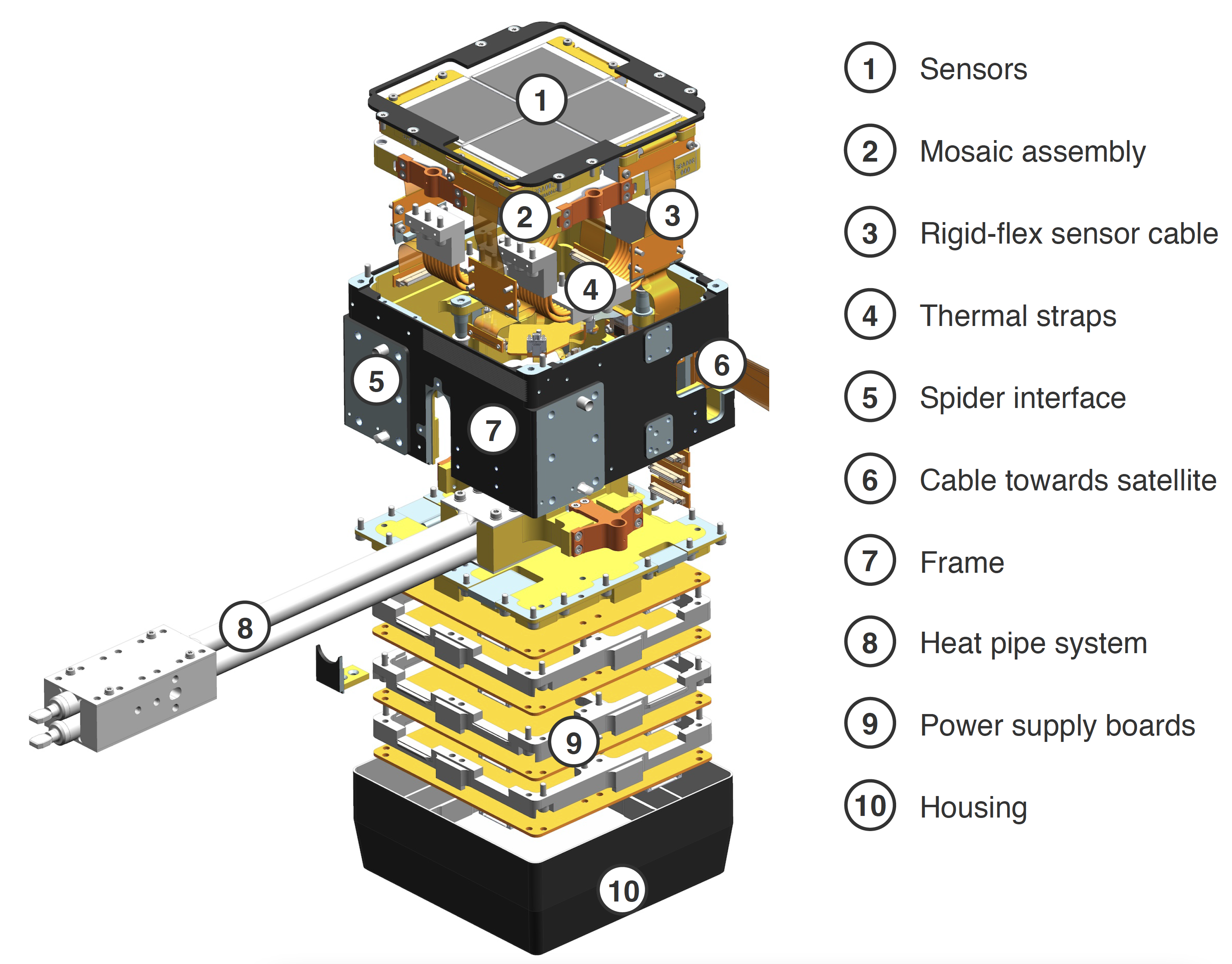} 
\end{tabular}
\end{center}
\caption{An exploded view of the ULTRASAT focal plane array and detector assembly. The ULTRASAT focal plane is comprised of four CMOS sensors aligned in a windmill configuration. The sensors are back illuminated devices with 22.4 megapixels and a sensitive area of $4.5\times4.5\,$cm$^2$ each. The DA houses the sensors and mosaic assembly, thermal interfaces and vicinity electronics boards. } 
\label{Fig:ULTRASAT_DA}
\end{figure}

\begin{figure}
\begin{center}
\begin{tabular}{cc}
\includegraphics[height=5cm]{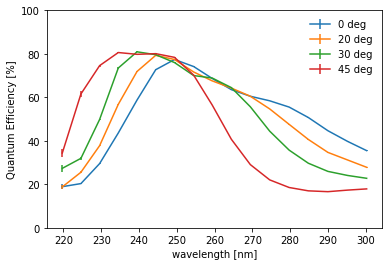} &
\includegraphics[height=5cm]{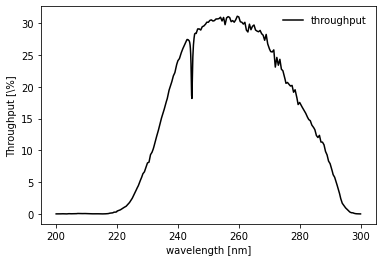}
\end{tabular}
\end{center}
\caption{Left: Measured quantum efficiency of a prototype scout sensor with ULTRASAT custom ARC\cite{2021SPIE11819E..0FB}; Right: Overall system efficiency, including optics, detector QE, and geometrical obscuration, averaged across the ULTRASAT FoV.} 
\label{Fig:ULTRASAT_Eff}
\end{figure}

At the heart of the DA is the mosaic assembly comprised of four packaged sensors mounted on a carrier plate, see Figure \ref{Fig:ULTRASAT_DA}. The sensors are complementary backside illuminated (BSI) metal-oxide-semiconductor(CMOS) devices with $22.4\,$megapixels and a sensitive area of $4.5\times4.5$cm$^2$ each. The sensors are passivated using a dielectric layer of Al$_2$O$_3$ with negative fixed charge. High quantum efficiency (QE), average of $59\%$ across the ULTRASAT band with a peak of $80\%$, is achieved due in part to a custom anti-reflective coating (ARC), taking into account both bandpass and incident angles of rays on the focal plane, see Figure \ref{Fig:ULTRASAT_Eff} left panel. Main performance parameters are shown in Table \ref{tab:detector performance}{}. The sensors were specifically designed and produced for the ULTRASAT mission by Analog Value Ltd. (AV) and Tower Semiconductor Ltd. (TSL).    

\begin{table}[]
    \centering
    \begin{tabular}{|l|c|c|}
        \hline
 Property & Value & Comment \\ [0.5ex] 
 \hline\hline
        Pixel Type & 5T dual gain & Rolling shutter configuration\\ \hline
        Pixel scale & $5.45\,$arcsec/pixel & $9.5\mu$m pixel with $f=360mm$ OTA\\ \hline
        Wavelength  & $230-290\,$nm & \\ \hline
        Quantum Efficiency & peak $>80\%$ & \\ \hline
        Dark Current @ $-73^{\circ}$C & $<0.026\,$e/pixel/sec & \\ \hline
        Readout Noise  & $<3.5\,$e& \\ \hline
    \end{tabular}
    \caption{Key performance parameters of the ULTRASAT detectors. Further information can be found in proceeding $12181-109$ of this volume.}
    \label{tab:detector performance}
\end{table}

The four CMOS sensors are aligned in a windmill configuration. Each sensor is bonded to a CE6 (a controlled expansion alloy of silicon and aluminium) carrier package using an epoxy adhesive, which allows to individually control out-of-plane translation and rotation to make the four tiles of the mosaic co-planar, as a high degree of flatness ($20\mu$m) is required for such a fast telescope. The sensors are wire bonded to a rigid-flex PCB assembly that additionally carries passive electronic components for noise filtering and connects each sensor to its individual power supply board. The mosaic assembly is thermally isolated from the frame by four flexures made of Ultem to allow for a temperature difference of around $80$K between the two elements.

Cooling of the mosaic assembly is realised by a thermal link between the sensor tiles and a thermal interface provided by the spacecraft. It is comprised of four copper rope thermal straps and two propylene heat pipes. Each thermal strap is connected to one of the sensor tiles to mechanically decouple the mosaic from the heat pipes. Temperature control of the sensors is performed by the spacecraft’s on-board computer with the help of heaters located on the sensor tiles. To achieve the higher temperatures required for decontamination of the sensors additional heaters are placed along the spacecraft's side of the thermal chain.

The DA‘s structural design is based on a milled frame, which ensures mechanical integrity and thus low displacements when being loaded mechanically or thermally. The frame is machined from titanium and provides the mechanical interfaces towards four “spiders” that connect the DA to the telescope. Furthermore, the frame provides the mechanical interfaces to a Sapphire filter window placed $550\mu$m above the sensors and a motorised focus mechanism that is attached at the side of the DA and allows for adjustments of the distance between first field flattener in the optical train and the mirror. The frame also provides interfaces to the interior of the DA, where the mosaic assembly, as well as the heat pipe system are attached. At the back of the frame a PCB stack accommodating the high precision power supplies for the sensors is located. An exploded view of the DA’s interior is shown in Figure \ref{Fig:ULTRASAT_DA}.

\subsection{Baffle}
\begin{figure}
\begin{center}
\begin{tabular}{cc}
\includegraphics[height=6cm]{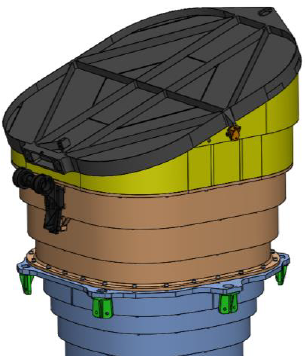} &
\includegraphics[height=6cm]{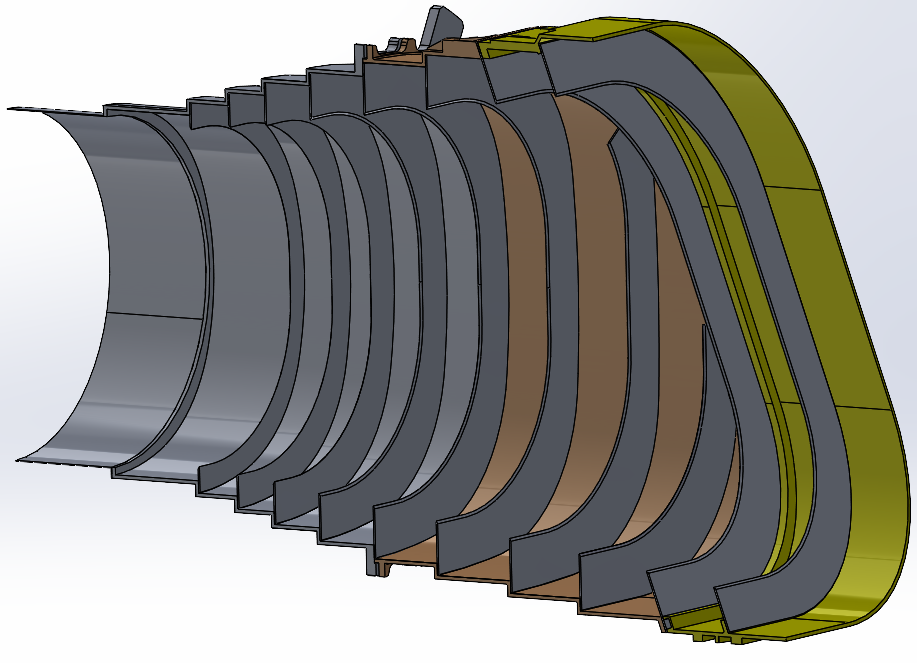}
\end{tabular}
\end{center}
\caption{Left: ULTRASAT baffle with the protective trap door closed; Right: A cross section of the baffle showing internal vane structure. The inside is coated with Acktar vacuum black.} 
\label{Fig:ULTRASAT_Baffle}
\end{figure}

A large baffle is placed in front of the telescope aperture stop. It is designed to suppress stray light into the OTA, as well as reduce the electron flux on the external corrector plane. It is fabricated from $4.2\,$mm thick aluminum sheets, and its interior is coated with Acktar vacuum black. At the edge of the baffle, a trap door is located to shield and protect the OTA during launch and transit to GEO. The Baffle is shown in Figure \ref{Fig:ULTRASAT_Baffle}. Further information can be found in section \ref{sec:SL}.

\section{Unique challenges and mitigation strategies}
\subsection{Radiation Environment and Cherenkov Radiation Background}
The harsh environment at GEO entails a significant radiation flux intercepting the OTA, specifically the front corrector lens\footnote{Radiation hardening of the detector is discussed in proceedings $12181-109$ of this volume, 'The design of a UV CMOS sensor for the ULTRASAT space telescope', T. Liran et al.}. While the substrate of the corrector lens is Corning HPF7980, a high purity synthetic fused silica with high transmission in the UV, radiation damage is known to create absorption centers, specifically in the UV, due to \textit{e.g.}, compactness, and generation of vacancies in the crystal structure. We therefore embarked on a dedicated campaign to both simulate the radiation environment on critical elements in the OTA and to measure the induced absorption at such levels.  

\begin{figure}
\begin{center}
\begin{tabular}{cc}
\includegraphics[height=6cm]{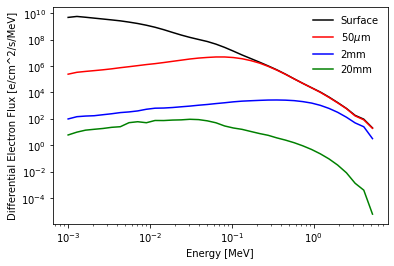} &
\includegraphics[height=6.2cm]{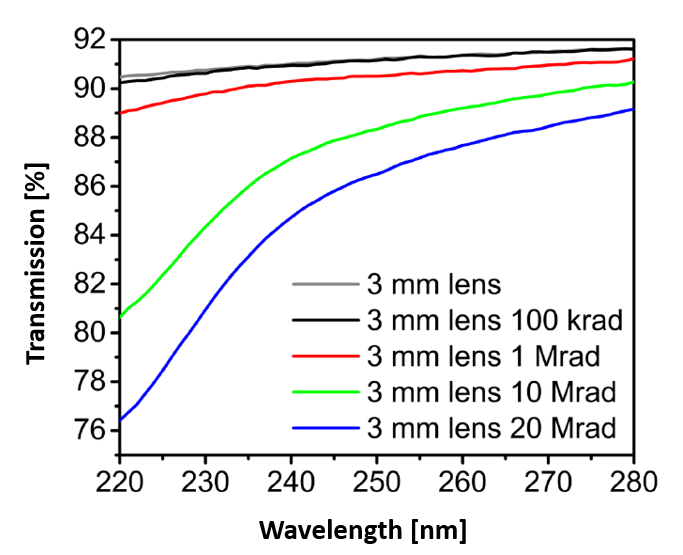}
\end{tabular}
\end{center}
\caption{Left: Simulated radiation level in the external corrector lens; Right: Transmission of witness samples of Corning HPF7980 before and after radiation, not corrected for Fresnel losses. We use these results to extrapolate, taking into account the difference in penetration length between the gamma radiation ($^{60}$Co) and energetic electrons (in orbit), to conclude an estimated transmission loss of $<2\%$ in a period of 3 years, and only at wavelengths shorter than $\sim245\,$nm.} 
\label{Fig:ULTRASAT_Radiation}
\end{figure}

Simulation of the radiation environment was performed by RASL (UK) based on the IGE2006 model. The external corrector outer surface receives a total ionising dose (TID) of $10^5$ Krads per year, dominated by trapped electrons in the orbit. The electron flux drops rapidly deeper into the lens, with an order of magnitude drop for electrons with energies of $\sim0.1\,$MeV at a depth of $\sim50\mu\,$m, see Figure \ref{Fig:ULTRASAT_Radiation} left panel. The expected electron dose above $0.1\,$MeV, the threshold for dislocations in the lattice, at the corrector surface is $\sim10\,$Mrad/yr, due in part to significant attenuation by the baffle. 

We then turned to measure UV transmission of the substrate samples at such radiation levels. Several samples of Corning HPF7980 were radiated using a $^{60}$Co source, for a total integrated radiation of up to $20\,$Mrad at Soreq NRC in Israel. UV transmission was measured before and after the samples were exposed to the radiation source using a dedicated setup and a monochromator. The results, see Figure \ref{Fig:ULTRASAT_Radiation} right panel, were then extrapolated to ULTRASAT based on the radiation analysis described above, and taking into account the different penetration depth of electrons and gamma radiation, the former being the dominant factor on orbit while the latter being the dominant factor for the $^{60}$Co source. We conclude expected transmission deterioration of $<2\%$ in a period of 3 years, and only at wavelengths shorter than $\sim245\,$nm.

The catadioptric design of the OTA, with refractive lenses at the front, raises another challenge due to Cherenkov radiation resulting from relativistic electrons propagating at velocities faster than the phase velocity in the substrate. This emission is predominantly in the UV, and will result in lower contrast and visibility in ULTRASAT images, similar effect to that of pupil ghosts. The baffle was designed to significantly reduce the electron flux on the corrector, and we find an estimated contribution of $0.1\,$photons$\,$s$^{-1}\,$ per pixel, which will result in noise contribution similar to that of the detector readout noise for integration time of $300\,$sec, and lower than the noise expected from zodaical light, see section \ref{sec:SL}.

\subsection{Out-of-Band Attenuation}
\label{sec:OoB}
The large ratio between the luminosity at visible band and at the UV band can reach more than 4 orders of magnitude for \textit{e.g.}, dM stars - the most abundant stellar objects in our local environment. Hence, a significant blocking capability at longer wavelengths is essential to obtain reliable data in the NUV band, and reduce background terms. The catadioptric nature of the optical system alleviates this requirement further, as the system PSF deteriorates significantly for $\lambda>300\,$nm. We therefore set a requirement on attenuation of radiation at $\lambda>290\,$nm of $<10^{-4}$.

\begin{figure}
\begin{center}
\begin{tabular}{cc}
\includegraphics[height=5.5cm]{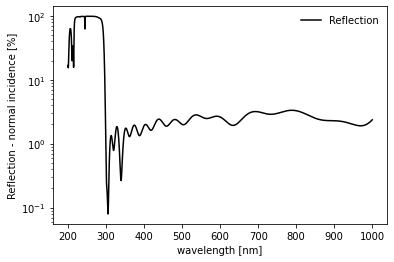} &
\includegraphics[height=5.5cm]{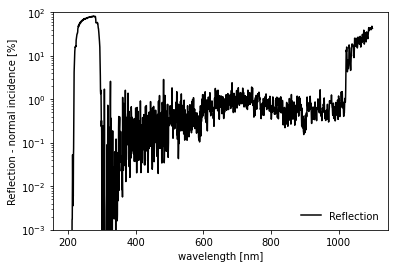}
\end{tabular}
\end{center}
\caption{Left: ULTRASAT mirror reflection. Out of band attenuation of $\sim1.4\%$ is achieved by the mirror ; Right: ULTRASAT Sapphire filter, developed and manufactured by VIAVI Solutions,  phase I measured data. We achieve more than two orders of magnitude attenuation at $\lambda>300\,$nm, allowing us to achieve our goal requirement of $<10^{-4}$ attenuation. Attenuation $>1000\,$nm is targeted in phase II.} 
\label{Fig:ULTRASAT_OoB}
\end{figure}

Out of band attenuation in the OTA is achieved in two steps: The use of a black mirror, and the implementation of a refractive filter in front of the detector. The former is realised by coating the spherical mirror with a custom coating of HfO and SiO$_2$ of $\sim30$ layers. The dielectric layers are designed such that radiation at the NUV bandpass is reflected by the mirror, while radiation at $\lambda>290\,$nm is transmitted through the mirror and absorbed by the mirror substrate and in light traps on the mirror back. The mirror achieves mean attenuation levels of $1.4\%$ for $\lambda>300\,$nm, see Figure \ref{Fig:ULTRASAT_OoB} left panel, and is designed, manufactured and tested by the OTA prime contractor.

The dielectric filter, located $0.55\,$mm in front of the focal plane array, is designed and manufactured by VIAVI Solutions. The attenuation is achieved by more than $1000$ layers of Al$_2$O$_3$/SiO$_2$ and HfO$_2$/SiO$_2$ coating stacks \cite{VIAVI}. A significant challenge is obtaining high transmission below $\sim250\,$nm. This is achieved by baking the filter at various stages in the coating process at temperatures of up to $700^{\circ}$C in order to fill in vacancies and imperfections at the boundaries within and between layers. Due to the high temperature required during the baking process, the filter substrate is Sapphire. In addition, the hardness of Sapphire allows us to set the filter thickness to only $4\,$mm while maintaining the stiffness required to resist stresses from the many layers of the coating. While Sapphire may have significant birefringence properties, this is mitigated by using Z-cut Sapphire substrate with the ordinary axis along the direction of the optical axis and placing the filter $0.55\,$mm in front of the detector. Measured transmission from a prototype Sapphire filter is shown in Figure \ref{Fig:ULTRASAT_OoB} right panel.

\subsection{Stray Light}
\label{sec:SL}
The low luminosity of UV sources requires us to achieve a suppression level of $\sim10^{-11}$ in the visible band for out-of-field sources in order to reach our target limiting magnitude of $>22$ in the ULTRASAT band. While some of the suppression is achieved by the filter and mirror coating described in the previous section, a significant effort must be placed on the design of a custom baffle and vane system to reach the required level of stray light suppression. The baffle is designed in collaboration with Breault engineering (USA) with emphasis on suppression of out-of-band light from the Earth, Sun, and Moon. The two main requirements for stray light suppression are:
\begin{itemize}
\item Earth total scattered light flux measured on the detector (both in-band and out of band) should result in $< 44,320\,$photons/cm$^{2}$/s, equivalent to irradiance suppression factor of $>4.5\times10^{10}$
\item Other sources (Moon, Sun diffracted by the baffle edge, etc.) should result in scattered light flux measured on the detector of $<10,000\,$photons/cm$^{2}$/s. This should be taken in light of our pointing requirement of half the celestial sphere available at any given moment.
\end{itemize}

The derived baffle design is shown in Figure \ref{Fig:ULTRASAT_Baffle}. Key features are: 
\begin{itemize}
    \item The baffle aperture has an angle of $20^{\circ}$, the longer side pointing towards the sun during standard observations. 
    \item Vane angles change at the entrance of the baffle, with no vane at the baffle exit port.
    \item The baffle interior is coated with Acktar vacuum black to minimize reflectance from baffle walls/vanes. The vane edge thickness is the minimum that can guarantee adequate coating of the knife edge ($\sim200\mu$m).
\end{itemize}

\section{Summary and Timeline}
\begin{figure}
\begin{center}
\includegraphics[height=8cm]{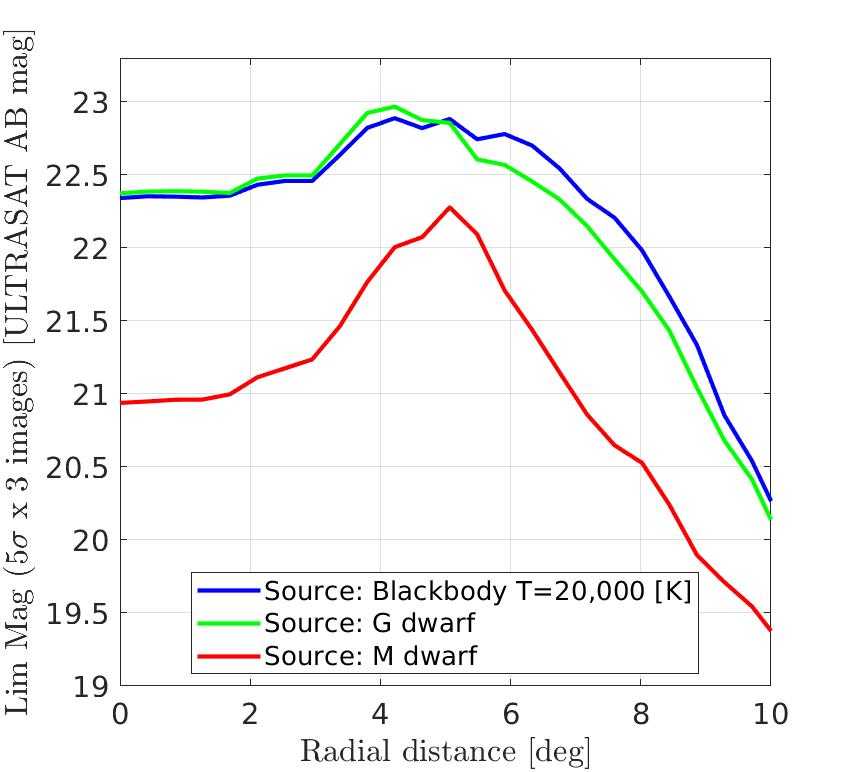}
\end{center}
\caption{ULTRASAT limiting magnitude. With its unprecedented efficiency, ULTRASAT achieves an average limiting magnitude of $22.4$ across a field of view of $>170\,$deg$^2$.} 
\label{Fig:ULTRASAT_LimMag}
\end{figure}

ULTRASAT is a unique UV telescope with an exceptional large field of view dedicated for the study of the transient sky. Its unique optical design allows us to achieve mean image quality of $\sim10\,$arcsec, while the choice of materials and its unique dielectric layer passivated sensor with custom ARC allow us to achieve efficiency of $\sim30\%$. The limiting magnitude for blackbody sources with $T>20,000$K is $\sim22.4$ for $3\times300\,$sec exposure, see Figure \ref{Fig:ULTRASAT_LimMag}. Compared to previous UV missions, ULTRASAT offers a grasp $300$ times larger than GALEX, the most sensitive UV satellite to date, and comparable to the upcoming Vera C. Rubin observatory. Its unique capabilities will allow us to detect EM radiation from GW sources, discover young CC-SNe during shock breakout and shock cooling phases, and determine which stellar hosts are more likely to harbor habitable planets.   

UV observations hold unique challenges, and this is further enhanced in the case of ULTRASAT given its unprecedented FoV and catadioptric design. We therefore adopt several unique solutions and mitigation techniques, among them a fine-tuned baffle to suppress stray light as well as the high energy electron flux in orbit, using refractive elements from high purity materials, and the use of unique multi-layer ($>1000$) dielectric coatings to suppress out-of-band flux. Further details regarding the design, challenges and verification and test plan will be detailed in future dedicated publications.  

In the past year, ULTRASAT passed the CDR stage for both the payload and bus. All critical and long lead items are now in hand or were ordered, and a payload MRR is expected to take place by the end of the year. This falls in line with the project plan to be launched during Q2/Q3 of 2025, and be fully operational during phase 5 of the large scale GW observatories.
\newpage
\bibliography{report} 

\begin{thebibliography}{10}

\bibitem{Ofek2020}
{Ofek}, E.~O. and {Ben-Ami}, S., ``{Seeing-limited Imaging Sky
  Surveys{\textemdash}Small versus Large Telescopes},'' {\em Publications of
  the Astronomical Society of the Pacific}~{\bf 132},  125004 (Dec. 2020).

\bibitem{LSST}
{Ivezi{\'c}}, {\v{Z}}., {Kahn}, S.~M., and {Tyson}, J.~Anthony, e.~a., ``{LSST:
  From Science Drivers to Reference Design and Anticipated Data Products},''
  {\em The Astrophysical Journal}~{\bf 873},  111 (Mar. 2019).

\bibitem{2016ARNPS..66...23F}
{Fern{\'a}ndez}, R. and {Metzger}, B.~D., ``{Electromagnetic Signatures of
  Neutron Star Mergers in the Advanced LIGO Era},'' {\em Annual Review of
  Nuclear and Particle Science}~{\bf 66},  23--45 (Oct. 2016).

\bibitem{2017ApJ...848L..32M}
{McCully}, C., {Hiramatsu}, D., and {Howell}, e.~a., ``{The Rapid Reddening and
  Featureless Optical Spectra of the Optical Counterpart of GW170817, AT
  2017gfo, during the First Four Days},'' {\em The Astrophysical Journall}~{\bf
  848},  L32 (Oct. 2017).

\bibitem{2018MNRAS.481.3423W}
{Waxman}, E., {Ofek}, E.~O., {Kushnir}, D., and {Gal-Yam}, A., ``{Constraints
  on the ejecta of the GW170817 neutron star merger from its electromagnetic
  emission},'' {\em Monthly Notices of the Royal Astronomical Society}~{\bf
  481},  3423--3441 (Dec. 2018).

\bibitem{2018LRR....21....3A}
{Abbott}, B.~P., {Abbott}, R., and {Abbott}, T.~D., e.~a., ``{Prospects for
  observing and localizing gravitational-wave transients with Advanced LIGO,
  Advanced Virgo and KAGRA},'' {\em Living Reviews in Relativity}~{\bf 21},  3
  (Apr. 2018).

\bibitem{2016ApJ...820...57G}
{Ganot}, N., {Gal-Yam}, A., and {Ofek}, Eran.~O., e.~a., ``{The Detection Rate
  of Early UV Emission from Supernovae: A Dedicated Galex/PTF Survey and
  Calibrated Theoretical Estimates},'' {\em The Astrophysical Journal}~{\bf
  820},  57 (Mar. 2016).

\bibitem{2011ApJ...728...63R}
{Rabinak}, I. and {Waxman}, E., ``{The Early UV/Optical Emission from
  Core-collapse Supernovae},'' {\em The Astrophysical Journal}~{\bf 728},  63
  (Feb. 2011).

\bibitem{2017ApJ...848....8R}
{Rubin}, A. and {Gal-Yam}, A., ``{Exploring the Efficacy and Limitations of
  Shock-cooling Models: New Analysis of Type II Supernovae Observed by the
  Kepler Mission},'' {\em The Astrophysical Journal}~{\bf 848},  8 (Oct. 2017).

\bibitem{2020ApJ...899...51S}
{Soumagnac}, M.~T., {Ofek}, E.~O., and {Liang}, Jingyi, e.~a., ``{Early
  Ultraviolet Observations of Type IIn Supernovae Constrain the Asphericity of
  Their Circumstellar Material},'' {\em The Astrophysical Journal}~{\bf 899},
  51 (Aug. 2020).

\bibitem{MUSCLES}
{France}, K., {Loyd}, R.~O.~P., and {Youngblood}, Allison, e.~a., ``{The
  MUSCLES Treasury Survey. I. Motivation and Overview},'' {\em The
  Astrophysical Journal}~{\bf 820},  89 (Apr. 2016).

\bibitem{HAZMAT}
{Shkolnik}, E.~L. and {Barman}, T.~S., ``{HAZMAT. I. The Evolution of Far-UV
  and Near-UV Emission from Early M Stars},'' {\em The Astronomical
  Journal}~{\bf 148},  64 (Oct. 2014).

\bibitem{Snellen}
{Snellen}, I.~A.~G., {de Kok}, R.~J., {le Poole}, R., {Brogi}, M., and
  {Birkby}, J., ``{Finding Extraterrestrial Life Using Ground-based
  High-dispersion Spectroscopy},'' {\em The Astrophysical Journal}~{\bf 764},
  182 (Feb. 2013).

\bibitem{Ben-Ami}
{Ben-Ami}, S., {L{\'o}pez-Morales}, M., {Garcia-Mejia}, J., {Gonzalez Abad},
  G., and {Szentgyorgyi}, A., ``{High-resolution Spectroscopy Using Fabry-Perot
  Interferometer Arrays: An Application to Searches for O2 in Exoplanetary
  Atmospheres},'' {\em The Astrophysical Journal}~{\bf 861},  79 (July 2018).

\bibitem{2018ApJ...864...27S}
{Stern}, D., {McKernan}, B., and {Graham}, Matthew~J., e.~a., ``{A Mid-IR
  Selected Changing-look Quasar and Physical Scenarios for Abrupt AGN
  Fading},'' {\em The Astrophysical Journal}~{\bf 864},  27 (Sept. 2018).

\bibitem{2015ApJ...806..129E}
{Edelson}, R., {Gelbord}, J.~M., and {Horne}, K., e.~a., ``{Space Telescope and
  Optical Reverberation Mapping Project. II. Swift and HST Reverberation
  Mapping of the Accretion Disk of NGC 5548},'' {\em The Astrophysical
  Journal}~{\bf 806},  129 (June 2015).

\bibitem{2013ApJ...779..187K}
{Kelly}, B.~C., {Treu}, T., {Malkan}, M., {Pancoast}, A., and {Woo}, J.-H.,
  ``{Active Galactic Nucleus Black Hole Mass Estimates in the Era of Time
  Domain Astronomy},'' {\em The Astrophysical Journal}~{\bf 779},  187 (Dec.
  2013).

\bibitem{2019ApJ...883...94T}
{Trakhtenbrot}, B., {Arcavi}, I., and {MacLeod}, Chelsea~L., e.~a., ``{1ES
  1927+654: An AGN Caught Changing Look on a Timescale of Months},'' {\em The
  Astrophysical Journal}~{\bf 883},  94 (Sept. 2019).

\bibitem{2019NatAs...3..242T}
{Trakhtenbrot}, B., {Arcavi}, I., and {Ricci}, Claudio, e.~a., ``{A new class
  of flares from accreting supermassive black holes},'' {\em Nature
  Astronomy}~{\bf 3},  242--250 (Jan. 2019).

\bibitem{2017ApJ...843..106B}
{Blanchard}, P.~K., {Nicholl}, M., {Berger}, E., {Guillochon}, J., {Margutti},
  R., {Chornock}, R., {Alexander}, K.~D., {Leja}, J., and {Drout}, M.~R.,
  ``{PS16dtm: A Tidal Disruption Event in a Narrow-line Seyfert 1 Galaxy},''
  {\em The Astrophysical Journal}~{\bf 843},  106 (July 2017).

\bibitem{2019MNRAS.489..524K}
{Kubota}, A. and {Done}, C., ``{Modelling the spectral energy distribution of
  super-Eddington quasars},'' {\em Monthly Notices of the Royal Astronomical
  Society}~{\bf 489},  524--533 (Oct. 2019).

\bibitem{2011blho.book..286G}
{Gezari}, S., ``{Tidal disruptions of stars by supermassive black holes},'' in
  [{\em Black Holes}{\nolinebreak\hspace{0.1em}]},  {Livio}, M. and
  {Koekemoer}, A.~M., eds.,  286--293 (2011).

\bibitem{2014ApJ...793...38A}
{Arcavi}, I., {Gal-Yam}, A., and {Sullivan}, Mark, e.~a., ``{A Continuum of H-
  to He-rich Tidal Disruption Candidates With a Preference for E+A Galaxies},''
  {\em The Astrophysical Journal}~{\bf 793},  38 (Sept. 2014).

\bibitem{2020SSRv..216..124V}
{van Velzen}, S., {Holoien}, T. W.~S., {Onori}, F., {Hung}, T., and {Arcavi},
  I., ``{Optical-Ultraviolet Tidal Disruption Events},'' {\em Space Science
  Reviews}~{\bf 216},  124 (Oct. 2020).

\bibitem{2020SSRv..216..114R}
{Roth}, N., {Rossi}, E.~M., {Krolik}, J., {Piran}, T., {Mockler}, B., and
  {Kasen}, D., ``{Radiative Emission Mechanisms},'' {\em Space Science
  Reviews}~{\bf 216},  114 (Oct. 2020).

\bibitem{2016MNRAS.461..371K}
{Kochanek}, C.~S., ``{Tidal disruption event demographics},'' {\em Monthly
  Notices of the Royal Astronomical Society}~{\bf 461},  371--384 (Sept. 2016).

\bibitem{2016ApJ...818L..21F}
{French}, K.~D., {Arcavi}, I., and {Zabludoff}, A., ``{Tidal Disruption Events
  Prefer Unusual Host Galaxies},'' {\em The Astrophysical Journall}~{\bf 818},
  L21 (Feb. 2016).

\bibitem{Schmidt}
{Wolfschmidt}, G., ``{The development of the Schmidt telescope},'' {\em
  Astronomische Nachrichten}~{\bf 330},  555 (June 2009).

\bibitem{2021SPIE11821E..0UA}
{Asif}, A. and {Barschke}, M.F., e.~a., ``{Design of the ULTRASAT UV camera},''
  in [{\em Society of Photo-Optical Instrumentation Engineers (SPIE) Conference
  Series}{\nolinebreak\hspace{0.1em}]},  {\em Society of Photo-Optical
  Instrumentation Engineers (SPIE) Conference Series} {\bf 11821},  118210U
  (Aug. 2021).

\bibitem{2021SPIE11819E..0FB}
{Bastian-Querner}, B., {Kaipachery}, N., and {K{\"u}ster}, Daniel, e.~a.,
  ``{Sensor characterization for the ULTRASAT space telescope},'' in [{\em
  Society of Photo-Optical Instrumentation Engineers (SPIE) Conference
  Series}{\nolinebreak\hspace{0.1em}]},  {\em Society of Photo-Optical
  Instrumentation Engineers (SPIE) Conference Series} {\bf 11819},  118190F
  (Aug. 2021).

\bibitem{VIAVI}
{Ockenfus}, G., ``{Metal-oxide Ultraviolet Narrow-band-pass Filter at 214
  nm},'' {\em Optical Interference Coatings}  (2013).

\end{thebibliography}
\bibliographystyle{spiebib} 

\end{document}